\documentstyle[11pt]{article}
\parskip = \baselineskip

\def\be{\begin{equation}}
\def\ee{\end{equation}}
\def\Tr{{\rm Tr~}}
\def\ov{\overline}
\def\conv{\displaystyle *}

\title{ \bf Entropy and Wigner Functions}
\author{G. Manfredi$^{1,*}$ and M. R. Feix$^2$}
\date{}
\begin{document}
\maketitle
\thispagestyle{empty}
\begin{center}
$^1${\em Laboratoire de Physique des Milieux Ionis\'es, Universit\'e
Henri  Poincar\'e, \\
BP239, 54506 Vandoeuvre-les-Nancy, France} \\
$^2${\em Subatech, Ecole des Mines de Nantes, BP 20722,
44307 Nantes Cedex 3, France} \\
\end{center}
$^{*}$ Electronic address : giovanni.manfredi@lpmi.uhp-nancy.fr

\begin{abstract}
The properties of an alternative
definition of quantum entropy, based on Wigner functions,
are discussed. Such definition emerges naturally from the Wigner
representation of quantum mechanics, and can easily quantify the amount
of entanglement of a quantum state. It is shown that smoothing of the Wigner function
induces an increase in entropy. This fact is used to derive some simple rules
to construct positive definite probability distributions which are also
admissible Wigner functions.
\end{abstract}


\vskip .5cm

{\large{\bf I. Introduction}}

Entropy is the central concept of thermodynamics and statistical
mechanics. It was introduced by Clausius in the mid-19th century
as a phenomenological variable that quantifies
the intrinsic irreversibilty of thermodynamical
processes. It was Boltzmann who recognized the link between entropy
and the lack of information about a system, defined as the number $\Gamma$
of microstates which have the same macroscopic properties.
The celebrated formula
\be
S_{\rm B} = k_{\rm B} \ln\Gamma
\label{sboltz}
\ee
establishes such a link in a mathematically rigorous manner
(in the rest of this article we shall use units for which
$k_{\rm B}=1$: with
this prescription, entropy becomes a dimensionless quantity).
Boltzmann, of course, derived this formula in the context of
classical statistical mechanics. In classical physics, microstates
are defined as points in a continuous $2D$-dimensional
phase space ($D$ is the number of degrees of freedom of the system
under consideration), and
cannot be "counted" in any meaningful sense. Therefore, Boltzmann
took as the number $\Gamma$ of microstates the available volume in phase
space $\Omega$ divided by the volume of a unit cell (unspecified at the time
when Boltzmann published his work, but which will
turn out to be Planck's constant,raised to the appropriate
power, $h^D$): $\Gamma=\Omega/h^D$.
In quantum mechanics, a microstate
is described by a wave function, which contains all
the information about the state of the system.
In contrast to the classical case, now
there is no ambiguity, since quantum states are discrete in principle.
Hence, although the macrostate has a huge number of possible microstates
consistent with it, this number, $\Gamma$, is nevertheless definite and finite.

The most general quantum system is described by a density matrix, i.e. a
positive-definite, Hermitian operator, with unit trace. In terms of the density
matrix $\rho$, the entropy can be expressed in the following way, due to
Von Neumann \cite{vn}
\be
S_{\rm VN} = -\Tr \rho \ln\rho .
\label{svn}
\ee
This is the standard definition of entropy, which generalizes Boltzmann's
expression to quantum mechanics. Although unambiguously defined, however,
$S_{\rm VN}$ can be extremely difficult to compute in practice, since
one would need to diagonalize $\rho$ in order to compute the trace of its
logarithm. Von Neumann's entropy (VN) has a number of good properties, which will
be detailed in the following sections. Here we note that,
if ${\alpha_i \ge 0}$ are the eigenvalues of the density matrix ($\sum_i \alpha_i=1$),
the VN entropy becomes $S_{\rm VN} = -\sum_i \alpha_i \ln\alpha_i$. Therefore
$S_{\rm VN}\geq 0$, and the equality holds only if we have complete information,
i.e. if only one of the eigenvalues
is different from zero: in this case, the system is in the pure state
corresponding to this eigenvalue.
Another crucial property of $S_{\rm VN}$ is that it is conserved as $\rho$
evolves according to the quantum Liouville equation
\be
i \hbar \frac{\partial\rho}{\partial t} = H\rho - \rho H~,
\label{qliouv}
\ee
where $H$ is the Hamiltonian. Indeed, the trace of any functional $F$
of the density matrix $\Tr F(\rho)$ is also conserved. This fact can
be used to define other entropy-like quantities.
Not all this quantities are equivalent, however, and we will
show in the following section that only one of them is particularly
adapted to the Wigner representation of quantum mechanics.

The classical limit of the Von Neumann entropy, Eq. (\ref{svn}), is obtained
by replacing the density matrix with the phase space
probability distribution $f(x,p)$ (for simplicity,
we will consider systems with only one degree of freedom, $D=1$),
and the trace with the integral in phase space. One obtains the
following expression, due to Gibbs
\be
S_{\rm CL} = - \int f\ln (fh) ~dx dp~,
\label{sclass}
\ee
and the probability distribution is positive and normalized to unity.
Note that the classical entropy is defined up to an additive constant,
which means that the constant $h$ in the argument of the logarithm in Eq. (\ref{sclass})
can be chosen arbitrarily, although it seems reasonable to use Planck's constant
$h=2\pi\hbar$. Indeed, if $f$ is constant inside a certain phase space volume $\Omega$
and zero elsewhere (i.e. at thermodynamic equilibrium),
then $S_{\rm CL}=\ln (\Omega/h)$, in agreement with Boltzmann's
original definition, Eq. (\ref{sboltz}).
We also stress that $S_{\rm CL}$ can take negative values, in contrast with
$S_{\rm VN}$, which is always non-negative. From the previous discussion, it is easy
to conclude that $S_{\rm CL}$ will be negative when $\Omega<h$.
This means that we are trying to localize a particle on a phase space
region smaller than Planck's constant, and therefore violate the uncertainty
principle. For probability distributions that satisfy the uncertainty principle,
the classical entropy is positive.
Similarly to the quantum mechanical case, the classical entropy
is conserved for a Hamiltonian process, i.e. when the probability distribution
evolves according
to the classical Liouville equation. Again, the phase space integral of any
functional $F(f)$ is also conserved (indeed, $f$ itself is conserved,
since it is just transported along the classical trajectories).

In this paper, we discuss the properties of an
alternative definition of quantum
entropy, based on Wigner functions.
Although this entropy has already been known for some time
(generally expressed in terms of the density matrix), we feel that its
properties are not fully appreciated. In particular,
it will be shown that such a definition of entropy emerges naturally
from the Wigner representation of quantum mechanics.
It has therefore a privileged status compared to
the many other definitions proposed in the literature,
and deserves to be studied in some depth.

The Wigner representation \cite{wig} is a useful tool to express
quantum mechanics in a phase space formalism (for reviews see
\cite{tat,hill}).
Although it was derived by Wigner for technical purposes, this
approach has recently attracted much interest, since it is  well-suited
to analyze the transition from classical to quantum dynamics.
The Wigner representation can deal with both pure and mixed quantum states,
and is completely equivalent to the more usual picture
based on the density matrix. In this representation, a
quantum state is described by a Wigner function (i.e. a function of the phase
space variables $-$ see next Section), and the Wigner equation
provides an evolution equation for the state which is equivalent to the
quantum Liouville equation (\ref{qliouv}). It will be shown that,
if one tries to define an entropy functional in the framework
of Wigner's representation,
only one `reasonable' choice is possible, and this is discussed in the next
Section. Subsequently, we will discuss the properties of such an entropy (Sec. III),
and present some examples of its applications in Secs. IV and V.

{\large{\bf II. Quantum Entropy}}

The quantum distribution function $W(x,p)$ is defined
in terms of the density matrix $\rho(x,y)$ for a quantum mixed state
\be
W(x,p)=\frac{1}{2 \pi\hbar}\int \rho\left(x-\frac{\lambda}{2},
x+\frac{\lambda}{2}\right)\exp\left(\frac{ip\lambda}{\hbar}\right)
d\lambda~,
\label{wigfun}
\ee
or in terms of the wavefunction $\psi(x)$ for a pure state
\be
W(x,p)=\frac{1}{2 \pi\hbar}\int \psi\left(x-\frac{\lambda}{2}\right)
\psi^{\displaystyle *}
\left(x+\frac{\lambda}{2}\right)\exp\left(\frac{ip\lambda}{\hbar}\right)
d\lambda~.
\label{wigpure}
\ee
The function $W(x,p)$
possesses many of the properties of a phase space probability distribution:
it is real, normalized to unity, and, when integrated over $x$ or $p$, gives
the correct marginal distribution, e.g. $\int W dp=\rho(x,x)= $
spatial density. Furthermore, it can be used to compute averages of any dynamical
variable $A(x,p)$ : $\langle A \rangle =  \int WA~dx dp$. Note however that,
since some terms in $A(x,p)$ may not commute, it is necessary to establish a
non-ambiguous correspondence between classical variables and quantum
operators (Weyl's rule) \cite{hill}.
Despite these good properties, the Wigner function cannot
be interpreted as a probability distribution,
since it can assume negative values. The only pure state whose Wigner
function is positive definite is given by the minimum uncertainty packet
(i.e. a Gaussian wavefunction).

The evolution of $W(x,p,t)$ is governed by the Wigner equation, which
replaces the classical Liouville equation :
\be
\frac{\partial W}{\partial t} + \frac{p}{m} \frac{\partial W}{\partial x} =
  \frac{i }{2 \pi
\hbar^2} \int \left[ \Phi \left(x - \frac{z}{2} \right) - \Phi \left(
x  + \frac{z}{2} \right) \right] \exp \left( - \frac{i}{\hbar}  (p -
p^\prime) z \right) W (x, p^\prime, t)  d z dp^\prime~,
\label{wigeq}
\ee
where $\Phi(x)$ is the  potential.
The Wigner equation is equivalent to the quantum Liouville equation (\ref{qliouv}),
and can describe the evolution of both pure states and mixtures.
However, in the present work, we shall privilege the Wigner formalism over
the density matrix one, since it is easier to represent in the classical phase space,
and it allows a more staightforward treatment of the semi-classical limit.

We would like to define an entropy functional in terms of Wigner functions.
The classical choice, Eq. (\ref{sclass}), obviously cannot work,
since $W$ can assume negative values.
It is easy to show the existence of two simple functionals of $W$ that
are invariant under Eq. (\ref{wigeq}): the first is the total probability
$\int W dx~dp =1$; the second invariant is $\int W^2 dx~dp$, which has no
obvious physical meaning.  We stress that this is a property of Eq. (\ref{wigeq}),
and does not depend on whether $W$ represents a pure state, a mixture, or even
a state which violates the uncertainty principle.
However, the fact that the latter expression is indeed invariant,
suggests that we introduce the following definition of entropy

\be
S_2 = 1 - (2\pi \hbar)^D \int W^2 dx~dp~,
\label{s2}
\ee

where $D$ is the number of degrees of freedom: except where otherwise stated, we will
always work with systems for which $D=1$.

The $S_2$ entropy can be expressed in
terms of the density matrix $\rho$
\be
S_2 = 1- \Tr \rho^2 ~,
\label{s2r}
\ee
a result which follows from the fact that $W$ is related to the Fourier transform
of $\rho$.  Equation (\ref{s2r}) has been used in the literature
as an entropy-like quantity \cite{zhp}, and sometimes referred to as the
linear entropy. Its relevance to Wigner functions has been noticed by some authors
\cite{pat}, but its full implications have not, to our knowledge, been appreciated and
developed. We first notice that
this is the only expression of entropy having the same functional
form when expressed in terms of either $W$ or $\rho$ (for example, $\int W^4$ is {\it not}
simply related to $\Tr \rho^4$). Secondly, and most importantly, the very structure
of Wigner's equation selects the functional $S_2$ as a special candidate for a
definition of entropy. It is therefore important to study its properties and implications.

When $W$ is an admissible Wigner function (i.e. when it represents either a pure
or a mixed quantum state), the previous entropy satisfies the relation
$0 \le S_2 \le 1$, and $S_2=0$ holds for a pure state,
which is a reasonable result, since
pure states contain the maximum information available.
Indeed, it is possible to define {\it quantum information} as the complement
of $S_2$ to unity, $I = 1-S_2$.
Note that $S_2$ can become negative only
for states that violate the uncertainty principle, as it will be explained in Sec. III.
We point out that $S_2=0$ is a necessary, but definitely
not sufficient condition for the corresponding Wigner function to represent a
pure state \cite{tat}. This can be shown by finding a counter-example. Let
us define the Wigner function as $W = \sum_{i=1}^3 \alpha_i W_i$,
where the $W_i$ are orthogonal pure states, and $\alpha_1=\alpha_2=2/3$,
$\alpha_3=-1/3$. Even though the coefficients $\alpha_i$ sum up to unity,
$W$ does not represent an admissible Wigner function, since one of the coefficients
(which represent probabilities) is negative. However, it is simple to prove
that $S_2[W]=0$.
Incidentally, this example has shown the existence of phase space functions
which represent neither pure states nor mixtures. This point will be discussed in more
detail in the next Section.

This entropy is related to a formula proposed by Tsallis \cite{tsa},
which has stimulated much work in the last decade (see, for example,
\cite{bogh} and references therein). If $\{\alpha_i\}$
is a set of probabilities adding up to unity, Tsallis entropy is defined by
\be
S_q =  \frac{1-\sum_i \alpha_i^q}{q-1}~,
\label{stsallis}
\ee
where $q$ is a real, not necessarily positive, number, and the standard entropy
is recovered for $q\to 1$. Tsallis entropy is a possible, and indeed useful, way to
generalize the  Boltzmann-Von Neumann expression,
and has been employed by several authors to study the thermodynamics of
strongly correlated systems, such as self-gravitating gases and inviscid
fluids \cite{bogh}.

Equation (\ref{s2}), is the continuous counterpart
of the discrete Tsallis entropy with $q=2$.
The continuous formula can be recovered by the following heuristic
argument. Let us cover the phase space
with cells of size $\Delta x \Delta p$. The discrete probabilities are then
$\alpha_i= W(x_i,p_i)\Delta x \Delta p$, and the discrete entropy becomes
\be
S_2 = 1-\Delta x \Delta p \sum_i W^2(x_i,p_i) \Delta x \Delta p~.
\label{sdiscr}
\ee
The sum in Eq. (\ref{sdiscr}) gives the integral $\int W^2 dx dp$. However,
we cannot let the factor $\Delta x \Delta p$ in front of the sum go to zero,
since this would violate the uncertainty relation. Indeed, we obtain the
correct continuous formula [Eq. (\ref{s2}) with $D=1$]
by taking for $\Delta x \Delta p$ the smallest value allowed
by quantum mechanics, i.e. Planck's constant $h=2\pi\hbar$.

Another way to go from the continuous to the discrete formula, is to
consider a Wigner function that is the sum of $N$ orthogonal pure states
$W(x,p) = \sum_{i=1}^N \alpha_i W_i(x,p)$. Of course $W$ represents a quantum
mixture. We recall the following useful relation, valid for orthogonal pure states:
\be
\int W_i W_j~ dx~dp = \delta_{ij}/2\pi\hbar~,
\label{scal}
\ee
where $\delta_{ij}$ is the Kronecker delta.
By developing $W$ in terms of the $W_i$ in Eq. (\ref{s2}),
and making use of Eq. (\ref{scal}), we obtain Tsallis discrete
entropy $S_2 = 1-\sum_{i=1}^N  \alpha_i^2$.
We stress again that the above properties are valid for the quadratic
entropy $S_2$, but do not hold for other functionals involving higher powers
of $W$.

It is interesting to show that a local entropy $\sigma$
and an entropy flux $J_S$ can also
be defined:
\be
\sigma(x,t) = \int W dp - 2\pi\hbar \int W^2 dp~,~~~~~~
J_S(x,t) = \int {p \over m} W dp - 2\pi\hbar \int {p \over m} W^2 dp~.
\label{sloc}
\ee
Of course one has $S_2 = \int \sigma dx$. By multiplying Eq. (\ref{wigeq})
by $W$ and integrating over momentum space, one can prove that the
local entropy obeys a continuity equation :
\be
\frac{\partial \sigma}{\partial t} + \frac{\partial J_S}{\partial x} = 0~,
\label{cont}
\ee
which shows that entropy can be transfered from one spatial location
to another, but is globally conserved. The physical meaning of $\sigma$
is easier to grasp if we express it in terms of the density matrix
in the position representation. With the help of Eq. (\ref{wigfun})
one finds (we drop the time dependence)
\be
\sigma(x) =  \rho(x,x) - \int \vline ~\rho
(x- \lambda/2, x+\lambda/ 2)~\vline^{~2}
~d\lambda~.
\label{slocr}
\ee
Equation (\ref{slocr}) shows that entropy is closely related to the
off-diagonal terms of the density matrix. For a pure state,
$\rho(x,y) = \psi(x) \psi^\star(y)$ ($\psi$  is the wavefunction),
and the local entropy can be expressed in terms of the spatial density
$n(x)=|\psi(x)|^2 = \rho(x,x)$
\be
\sigma(x) = n(x)-\int n\left(x-{\lambda \over 2}\right)
n\left(x+{\lambda \over 2}\right)~d\lambda
\equiv n(x) - \iota(x)~,
\label{slocn}
\ee
where we have defined the {\it local quantum information}
$\iota(x)$ so that $I=\int \iota dx$.
It appears that $\iota(x)$ is a density autocorrelation function, which
shows that, in quantum mechanics, information and spatial correlations
are intimately close concepts.

{\large{\bf III. Properties of Quantum Entropy}}

The expression given in
Eq. (\ref{s2}) has proven to be a fruitful tool to quantify some key
properties of quantum systems, such as nonlocal correlations.
In order to be an appropriate definition of entropy it should nevertheless
satisfy some standard properties \cite{wehrl}, among which concavity and additivity are
particularly fundamental. Some of these properties were previously studied by
Tsallis \cite{tsa} for the discrete case.

{\it 1. Concavity.} This means that, if $W = \sum_{i=1}^N \alpha_i W_i$ (where the
$W_i$ are not necessarily pure orthogonal states), then the following inequality
holds
\be
S_2[W] \geq \sum_{i=1}^N \alpha_i S_2[W_i]~.
\label{conc}
\ee
The proof is obtained by direct calculation for $N=2$, and then is easily
extended to higher $N$ by recursive arguments.

Note that we can also prove an upper bound for $S_2$
\be
S_2[W] \leq \sum_{i=1}^N \alpha_i^2 S_2[W_i] + 1-\sum_{i=1}^N \alpha_i^2~,
\label{ubound}
\ee
which holds for $W_i$ representing both pure states or mixtures.
The term $1-\sum_i \alpha_i^2$ represent the so-called
mixing entropy.
The proof of Eq. (\ref{ubound}) relies on the following inequality \cite{tat}
\be
\int W_i W_j~ dx dp \ge 0~,
\label{ineq}
\ee
which is valid for all admissible Wigner functions, pure or mixed states
(see Sec. IV for a definition of admissibility).
When the $W_i$ represent
pure states, then $S_2[W_i]=0$, and Eq. (\ref{ubound}) becomes
\be
S_2[W] \leq 1-\sum_{i=1}^N \alpha_i^2~.
\label{uboundp}
\ee
The equality sign holds when the $W_i$ are also orthogonal, as was shown
in Sec. II.

{\it 2. Additivity.} Let us consider two independent subsystems
$A$ and $B$. The Wigner function $W$ describing the total system $A \cup B$
is simply given by the product of the Wigner functions $W_A$ and $W_B$ for
the two subsystems
\be
W(x_A,p_A,x_B,p_B) = W_A(x_A,p_A)~ W_B(x_B,p_B)~.
\label{wab}
\ee
It is easy to show that both the classical entropy, Eq. (\ref{sclass}), and
the Von Neumann entropy, Eq. (\ref{svn}), are additive \cite{wehrl}, i.e.
$S[W] = S[W_A]+S[W_B]$. This is a key property, since it enables one to
identify the statistical entropy with the thermodynamical entropy, which is
also additive.

By contrast, our definition of entropy is not additive in the usual sense.
Let us first notice that, whereas the number of degrees of freedom of each
subsystem is $D=1$, the total system has $D=2$. Therefore the information is defined
as $I[W_{A,B}]=h\int W_{A,B}^2$ for each subsystem, and  $I[W]=h^2\int W^2$ for the total
system. With this in mind, it is easy to establish the following expression for
the quantum information
\be
I[W] = I[W_A] ~I[W_B]~,
\label{iadd}
\ee
which shows that, since $I<1$, the information contained in the total system
is smaller than the information of each subsystem, except for pure states, for
which $I=1$. In terms of the entropy $S_2=1-I$, Eq. (\ref{iadd}) becomes
\be
S_2[W] = S_2[W_A] + S_2[W_B] - S_2[W_A]~ S_2[W_B] ~.
\label{sadd}
\ee
The total entropy is therefore smaller than the sum of the partial entropies,
but larger than each of them. Note that when the subsystems are "almost pure"
quantum states, then $S_2[W_{A,B}] \ll 1$, and the non-additive correction to
Eq. (\ref{sadd}) becomes of higher order. In this case, approximate additivity
is recovered.

It is also interesting to note that Eq. (\ref{sadd}) is formally identical to the
expression for the probability of the union of two subsets $A$ and $B$, which reads
\be
prob(A \cup B) = prob(A)+prob(B) - prob(A \cap B)~,
\label{prob}
\ee
and $prob(A \cap B) = prob(A)~prob(B)$ for statistically independent systems.
The analogy of $S_2$ as probability is also consistent with the
normalization $0 \le S_2 \le 1$.

{\it 3. Subadditivity.} If the subsystems $A$ and $B$ are not independent, the
Wigner function cannot be factored as in Eq. (\ref{wab}). The Wigner function of each
subsystem is then defined by integrating over the other system's variables,
for instance
\be
W_A(x_A,p_A) = \int W(x_A,p_A,x_B,p_B) dx_B dp_B~,
\ee
and similarly for $W_B$. For the
Boltzmann-Von Neumann entropy, one can prove that $S[W] \le S[W_A]+S[W_B]$,
and the equality sign holds when the two subsystems are independent
\cite{wehrl}. This means
that the total system $A \cup B$ contains {\it more} information than the sum of its
parts $-$ which is natural, since the two subsystems are correlated.
However, no such relation can be proven for $S_2$ : this entropy is therefore not
subadditive. Note that this fact is consistent with the analogy of $S_2$
as probability given by Eq. (\ref{prob}). Indeed, when the subsets $A$ and $B$
are not independent, the probability of their intersection $prob(A \cap B)$ can be
either smaller or larger than the product $prob(A)~prob(B)$, corresponding to
either negative or positive correlation.

{\it 4. Microcanonical Ensemble.} We want to extremize the entropy $S_2$ with the
constraint $\int W dx dp=1$. Using Lagrange multipliers, it is easy to show that the
entropy is maximum when $W={\rm const.} = \Omega^{-1}$ within a phase space region
of volume (area) equal to $\Omega$, and $W=0$ elsewhere. In this case the entropy is
\be
S_2 = 1 - \frac{h}{\Omega}~,~~~~(h=2\pi\hbar).
\label{smicr}
\ee
This is the analog of Boltzmann's formula, Eq. (\ref{sboltz}), when the appropriate
additive constant is used, i.e. $S_{\rm B} = \ln (\Omega/h)$. For both expressions,
$S=0$ when $\Omega=h$ (minimum uncertainty), and the entropy becomes negative
when $\Omega<h$, i.e. when the uncertainty relation is violated. In limit
$\Omega \rightarrow \infty$, $S_2$ is bounded, and tends to unity (least information).
With this notation, information $I=1-S_2$ is just the inverse of the number of
available microstates $\Omega/h$.

{\it 5. Canonical Ensemble.} We now extremize $S_2$ with the constraints $\int W dx dp=1$
and $\int W E dx dp=U$, where $E(x,p)=p^2/2m+\Phi(x)$, and $U$ is the average energy.
Again using Lagrange multipliers, we find the following equilibrium distribution
\be
\begin{array}{ll}
W_{\rm eq}(x,p)= Z^{-1} [1-\beta E(x,p)]~,& ~\beta E<1 \\
W_{\rm eq}(x,p)=0 ~,& ~\beta E \ge 1
\end{array}
\label{wcan}
\ee
where $\beta$ is the Lagrange multiplier corresponding to the energy constraint, and can
be interpreted in the usual fashion as the inverse temperature $\beta=1/T$; $Z$ is a
normalization constant. For energies such that $\beta E \ll 1$, Eq. (\ref{wcan})
becomes identical with the standard exponential Boltzmann factor $\exp(-\beta E)$.
Since $W_{\rm eq}$ is a linear function of the energy, we have been forced to introduce a
cut-off, otherwise $W_{\rm eq}$  would diverge for large values of $E$.
Physically, this means
that states with energy $E>T$ are forbidden at equilibrium. Note the difference with
standard thermodynamics, where such states are highly improbable (because Boltzmann's
factor decreases exponentially), but not forbidden in principle.

An interesting fact is that Eq. (\ref{wcan}) is a stationary solution of
the Wigner equation (\ref{wigeq}) $-$ indeed, we are aware of no other
stationary solution which is also a function of the energy $E(x,p)$ alone.
This is easy to prove when the right-hand side
of Eq. (\ref{wigeq}) is written as
\[
\sum_{n=0}^\infty c_n \frac{\partial^{2n+1} \Phi}{\partial x^{2n+1}}
\frac{\partial^{2n+1} W}{\partial p^{2n+1}}~,
\]
where the $c_n$ are constants.
The $n=0$ term yields the classical part of Wigner's equation, whereas all other
terms do not provide any contribution, since $W_{\rm eq}$ is quadratic in $p$.
Moreover, since $W_{\rm eq}$ is a
function of the energy alone, it is a stationary solution of the classical Liouville
equation, so that we have finally $\partial W_{\rm eq}/ \partial t = 0$.
The fact that maximizing the entropy $S_2$ naturally yields a Wigner function
which is both stationary and a function of the energy alone is in itself
remarkable. At the present stage, it is premature to make any statement about
the role of $W_{\rm eq}$, but the subject certainly deserves further attention.
For example, it would be interesting to know if, and under what constraints,
$W_{\rm eq}$ can act as an attractor in a relaxation process.

{\large{\bf IV. Smoothed Wigner Functions}}

The Wigner function cannot be interpreted as a genuine probability distribution
because it almost always takes negative values. The only pure state whose
Wigner function is positive is given by the minimum uncertainty Gaussian
wavepacket :
\be
\psi(x) = (2\pi)^{-1/4} \sigma^{-1/2} \exp(-x^2/4\sigma^2)~,
\label{gauss}
\ee
whose Wigner function is also Gaussian
\be
G(x,p) = \frac{1}{\pi \hbar} \exp\left(-\frac{x^2}{2\sigma^2}
-\frac{2 p^2\sigma^2}{\hbar^2} \right)~.
\label{wgauss}
\ee
A possible way to obtain a positive distribution is to
smooth a pure Wigner function $W(x,p)$ using a kernel $K(x,p)$
which is itself a Wigner function correponding to a
pure state \cite{smooth}.
The smoothing operation is represented mathematically by a
convolution in phase space. The smoothed Wigner function $\ov{W}(x,p)$
\be
\ov{W}(x,p)= \int W(x^\prime,p') K(x-x',p-p')~dx' dp'
\equiv W \conv K~,
\label{conv}
\ee
is then positive and normalized to unity, so that it can be interpreted as
a probability distribution.

In the past, the most common choice of the smoothing kernel
has been the minimum uncertainty Gaussian $G(x,p)$, as given in
Eq. (\ref{wgauss}) \cite{smooth}. The resulting smoothed Wigner
function is sometimes referred to as the Husimi function.
This choice is however quite arbitrary, and no argument has ever been
proposed, to our knowledge, in order to justify its privileged status.
We shall now prove that smoothing with a Gaussian kernel {\it does} have some
special properties, and should therefore be regarded as the correct way
to obtain positive smoothed Wigner functions. In particular,
it will be shown that, when the smoothing is performed with a Gaussian
kernel, {\it the result is still an admissible Wigner function}.

First of all, we need a precise definition of an admissible Wigner function. Of course,
not all functions of the phase space variables are admissible: for example, those
functions which violate the uncertainty principle are clearly not admissible.
Functions that can be constructed by summing orthogonal pure states, such as
$W = \sum_i \alpha_i W_i$, are not admissible if some of the $\alpha_i$
are negative : this was the example analyzed in Sec. IV.
Our definition of an admissible Wigner function is rather standard
\cite{hill}, and is based
on the density matrix formalism. According to standard quantum theory, a
density matrix $\rho$ must satisfy three properties in order to describe a
quantum mixed state : (1) it must have unit trace $\Tr \rho = 1$; (2) it must be Hermitian
$\rho(x,y)=\rho^{\displaystyle *}(y,x)$; and (3) its eigenvalues must be non-negative.
While the first two properties are easy to verify, the third is much harder to test,
since one would need to diagonalize $\rho$ in order to compute its eigenvalues.
Property (3) can also be expressed in the following way :
\be
\int \psi(x)~\rho(x,y)~\psi^{\displaystyle *}(y) dx~dy \ge 0,~~
\forall \psi,
\label{posdef}
\ee
where the inequality must hold for {\it all} wavefunctions $\psi$.
This makes it even more apparent that Property (3) cannot be used as an
operational test.

Now, the previous properties can be transposed to Wigner functions by
making use of the definition, Eq. (\ref{wigfun}). In particular we would like
to know whether the {\it smoothed} Wigner function $\ov W$ is in general
admissible or not. Properties (1) and (2) simply
require that $\ov W$ be real and normalized to unity. Property (3) can be written in
the following form \cite{hill}
\be
\int \ov W(x,p) F(x,p)~dx dp \ge 0,~~\forall F(x,p) = {\rm pure~state}~.
\label{posdefw}
\ee
The equivalence between Eqs. (\ref{posdef}) and (\ref{posdefw}) can be verified
by noting that $\ov W$ and $F$ are the Wigner transform of, respectively,
$\rho$ and $\psi$, as defined in Eqs. (\ref{wigfun}-\ref{wigpure}).
It is clear that, in order to check the admissibilty of $\ov W(x,p)$, one should perform
an infinite number of integrals involving test Wigner functions $F(x,p)$ that
represent pure states.
However, Eq. (\ref{posdefw}) can be used to prove that smoothing with a Gaussian
kernel yields a smoothed Wigner function which is itself admissible.

In order to do so, let us plug Eq. (\ref{conv}) into the left hand-side
of Eq. (\ref{posdefw}). We obtain ($W$ is the original Wigner function, $K$
is the smoothing kernel, and $F$ is the test function: all three represent pure states)
\be
\begin{array}{ll}
\int W(x-x',p-p') F(x,p) K(x',p')~dx' dp' dx dp \\
= \int K(x',p')~dx' dp' \int W_1(x-x',p-p') F(x,p)~dx dp \\
= \int K(x',p') [W_1 \conv F](x',p') ~dx' dp' ~,
\end{array}
\label{gsmooth}
\ee
where $W_1(x,p)=W(-x,-p)$ is the Wigner function corresponding to the
wavefunction $\psi(-x)$ [whereas $W$ corresponds to $\psi(x)$]. The term
$W_1 \conv F$ is certainly a positive function, since it is the
convolution product of two Wigner functions. It follows that a sufficient condition
for Eq. (\ref{posdefw}) to be satisfied is that $K(x,p)$ be positive. But the only
pure state Wigner function which is also positive is the Gaussian $G(x,p)$
[Eq. (\ref{wgauss})]. This proves that, when the smoothing kernel is Gaussian,
the inequality given in Eq. (\ref{posdefw}) is verified, and the smoothed Wigner function
$\ov W(x,p)$ is therefore admissible.
In this case, the density matrix $\ov \rho$
corresponding to $\ov W$ can be written as
\be
\ov \rho(x,y)=\frac{1}{\sqrt{2\pi}\sigma}\int W(q,p)\exp\left(-\frac{(x-q)^2}{4\sigma^2}
+\frac{ipx}{\hbar}\right)\exp\left(-\frac{(y-q)^2}{4\sigma^2}-\frac{ipy}{\hbar}\right)
dq dp~.
\label{coherent}
\ee
The previous result can be easily checked by computing the Wigner function
$\ov W$ associated to $\ov \rho$ via Eq. (\ref{wigfun}), and realizing that it
can be written as $\ov W = W \conv G$. Equation (\ref{coherent}) expresses the density
matrix as a continuous sum of localized states in phase space
(`coherent states' \cite{howard}). Note that the coefficients in this sum
[i.e. $W(x,p)$ itself] are not necessarily positive numbers. The reason for this is
that the set of coherent states is `overcomplete', meaning that the
representation of an arbitrary quantum state in terms of coherent states
is not unique. However, thanks to the previous
theorem, we know that a diagonal representation of $\ov \rho$ with
non-negative coefficients
does exist, although we are not generally able to construct it explicitly.

So far we have proven that smoothing with a Gaussian kernel yields a function $\ov W$
which is itself an admissible Wigner function. Nothing definite can be said when
the smoothing is performed using a different kernel. However, we are able to produce
a counterexample, i.e. a pure state Wigner function which, after smoothing with a
non-Gaussian kernel, does not satisfy Eq. (\ref{posdefw}), and is therefore not
admissible. Let us consider the wavefunction
\be
\psi(x) = 2 (2/\pi)^{1/4} x \exp(-x^2)~,
\label{example}
\ee
and call $W(x,p)$ its Wigner transform. Now we smooth $W$ using as kernel $W$ itself:
\be
\ov W = W \conv W~.
\ee
In order to be an admissible Wigner function, $\ov W$ must satisfy Eq. (\ref{posdefw})
for every test function $F$. Let us use as test function once again $W$ itself, and
compute the integral in Eq. (\ref{posdefw}). We obtain (details are in the Appendix)
\be
\int \ov W(x,p) W(x,p)~dx dp = -\frac{1}{27\pi\hbar} < 0~.
\label{counter}
\ee
This result shows that not all ways of smoothing Wigner functions are
equivalent: only by smoothing with a Gaussian kernel we are certain to obtain a function
which is positive and also represents an admissible quantum state (i.e. a state
defined by a density matrix with real non-negative eigenvalues).

Furthermore, Eq. (\ref{gsmooth}) suggests another way to construct a phase space
distribution which is both positive and admissible [satisfying Eq. (\ref{posdefw})].
Let us take for $W(x,p)$ an {\it arbitrary} positive function of phase space
variables, and smooth it with a Gaussian kernel $G(x,p)$: $\ov W = W \conv G$.
We want to prove that $\ov W$ is admissible. Equation (\ref{posdefw}) yields
(using the fact that $G$ is even)
\be
\begin{array}{ll}
\int W(x-x',p-p') F(x,p) G(x',p')~dx' dp' dx dp \\
= \int W(x',p')~dx' dp' \int G(x'-x,p'-p) F(x,p)~dx dp \\
= \int W(x',p') [G \conv F](x',p') ~dx' dp' > 0~.
\end{array}
\ee
The result follows from the fact that the convolution product is positive,
since both $F$ and $G$ are pure state Wigner functions, and $W>0$ because
we chose it to be so. This proves that $\ov W(x,p)$ is an admissible Wigner
function, and is also positive, since it is the convolution product of two
positive functions. The density matrix corresponding to $\ov W$ is again
$\ov \rho$, as given by Eq. (\ref{coherent}).
Physically, the smoothed function $\ov W = W \conv G$ can be interpreted as
the admissible quantum state which best approximates
the classical state $W$ for a given value of $\hbar$.

To conclude this Section, we restate the two main results that have been
obtained here. We have shown two possible ways to construct a phase space
distribution which is both positive and an admissible quantum state.
This can be performed ({\it a}) by smoothing a pure state Wigner function with a
Gaussian kernel, or ({\it b}) by smoothing an arbitrary (but positive) function of
phase space variables, again with a Gaussian kernel.
Therefore, the Gaussian function $G(x,p)$ given in Eq. (\ref{wgauss}) has a
privileged status as a smoothing kernel. Note, however, that $G$ is not
unique, since it depends on the parameter $\sigma$.

Although such results were derived for a pure state Wigner function, they can
easily be generalized to mixtures. It follows that, when smoothing
several times with a Gaussian kernel,
we still remain within the class of admissible
Wigner functions. This class is therefore closed with respect to this particular
operation.

{\large{\bf V. Entropy and Smoothed Wigner Functions}}

The smoothing operation has the effect of erasing some of the
correlations in the phase space. We expect therefore that smoothing
should increase the entropy. This is not difficult to prove. In order
to do this, we need to define the double Fourier transform of a Wigner
function $W(x,p)$
\be
W(k,\lambda) = \int \int W(x,p) \exp(-ikx-i\lambda p) dx~dp~.
\label{four}
\ee
By means of Eqs. (\ref{wigpure}) and (\ref{four}), one obtains for a
pure state
\be
W(k,\lambda) =  \int \psi \left(x-\frac{\lambda \hbar}{2}\right)
\psi^{\conv} \left(x+\frac{\lambda \hbar}{2}\right) \exp(-ikx) dx~.
\label{wigfour}
\ee
We can then easily prove the following Lemma
\be
|W(k,\lambda)|^2 \le~ \int{\vline\psi \left(x-\frac{\lambda \hbar}{2}\right)dx\vline~}^2~
\int{\vline\psi^{\conv} \left(x+\frac{\lambda \hbar}{2}\right)dx\vline~}^2 = 1~,
\label{lemma}
\ee
where use has been made of Schwartz's inequality.

Now, let us take an arbitrary Wigner function $W(x,p)$ and smooth it
with a kernel $K(x,p)$ which is a pure state: $\ov W(x,p) = W(x,p)\conv K(x,p)$.
In Fourier space we have : $\ov W(k,\lambda) = W(k,\lambda) K(k,\lambda)$.
The quantum information $I[\ov W] = 2\pi\hbar \int\ov W^2 dx dp$
relative to $\ov W$ satisfies the inequalities
\be
\begin{array}{ll}
I[\ov W] = \frac{\hbar}{2\pi}
\int |\ov W(k,\lambda)|^2 dk d\lambda
= \frac{\hbar}{2\pi} \int |W(k,\lambda)|^2 |K(k,\lambda)|^2 dk d\lambda \\
\le \max~ | K(k,\lambda)|^2 \times
\frac{\hbar}{2\pi} \int |W(k,\lambda)|^2 dk d\lambda \\
= \max ~| K(k,\lambda)|^2 \times 2\pi\hbar \int W^2 dx dp
\le 2\pi\hbar \int W^2 dx dp = I[W]~,
\end{array}
\label{entroinc}
\ee
where we have used the previous Lemma [Eq. (\ref{lemma})] for $K$,
as well as Parseval's identity in the form
\be
\int W^2(x,p)dx dp = \frac{1}{4\pi^2} \int |W(k,\lambda)|^2 dk d\lambda~.
\ee
Equation (\ref{entroinc}) implies that
\be
S_2[\ov W] \ge S_2[W]~,
\label{s_smooth}
\ee
i.e. the smoothing operation has increased the entropy. Note that, in order to obtain this
result, the smoothing kernel needs not be a Gaussian.

Now we turn to the case where the smoothing kernel is indeed Gaussian.
In this case, a relatively simple expression for $I[\ov W]$ can be obtained.
The double Fourier transform of the Gaussian defined in Eq. (\ref{wgauss}) is
\be
G(k,\lambda) = \exp\left(-\frac{k^2 \sigma^2}{2}
-\frac{\lambda^2 \hbar^2 }{8\sigma^2} \right)~.
\label{wgaussfour}
\ee
The Fourier transform of the Wigner function $W$ to be smoothed is
given by Eq. (\ref{wigfour}).
Let us compute the information :
\be
I[\ov W] = 2\pi\hbar\int \ov W^2 dx dp
= \frac{\hbar}{2\pi} \int |W(k,\lambda)|^2 |G(k,\lambda)|^2 dk d\lambda~.
\ee
Expressing $W$ and $G$ by means of Eqs. (\ref{wigfour}),(\ref{wgaussfour}) one obtains,
after some algebra
\[
I[\ov W] = \frac{1}{2 \sigma\sqrt{\pi}} \int \psi \left(x-\frac{\lambda}{2}\right)
\psi^{\conv} \left(x+\frac{\lambda}{2}\right) \psi \left(x'+\frac{\lambda}{2}\right)
\times
\]
\be
\psi^{\conv} \left(x'-\frac{\lambda}{2}\right)
\exp\left(-\frac{\lambda^2+(x-x')^2}{4\sigma^2} \right)dx dx' d\lambda~.
\ee
We now change the integration variables, using the following unitary
transformation
\be
\begin{array}{ll}
x  = & {1\over 2} w +{1\over 2} y +{1\over 2} z \\
x'  = & {1\over 2} w -{1\over 2} y - {1\over 2} z \\
\lambda  = & - y +  z ~.
\end{array}
\ee
After some algebra, the following result is obtained
\be
I[\ov W] = \frac{1}{\sigma\sqrt{\pi}} \int dw~\vline ~{\int \psi(w+y)\psi(w-y)
\exp\left(-{y^2 \over {2\sigma^2}}\right)dy~\vline~}^2~,
\label{infsmooth}
\ee
which expresses the quantum information in terms of the wavefunction $\psi$
corresponding to the unsmoothed Wigner function $W$. Equation (\ref{infsmooth})
may be usefully employed to monitor the time evolution of the entropy in a
numerical simulation: $\psi(x,t)$ would then evolve according to the time-dependent
Schr\"odinger equation.

Finally, we show that a more stringent bound than the one expressed by Eq.
(\ref{s_smooth}) can be obtained when the smoothing kernel is Gaussian.
By using again Schwartz's inequality, we have from Eq. (\ref{infsmooth})
\[
I[\ov W] \le \frac{1}{\sigma\sqrt{\pi}} \int dw \int dy
\exp\left(-{y^2 \over {\sigma^2}}\right) \int dy'~
{\vline~\psi(w+y')\psi(w-y')~\vline ~}^2
\]
\be
= \int dw \int dy'~{\vline~\psi(w+y')\psi(w-y')~\vline ~}^2
= {1 \over 2}~.
\ee
In terms of the entropy, this becomes
\be
S_2[\ov W] \ge {1 \over 2}~,
\label{s_smooth2}
\ee
a result that is valid when smoothing a pure Wigner function with a
Gaussian kernel.
Note that we still have some freedom in the choice of the kernel,
since the width $\sigma$ of the Gaussian in Eq. (\ref{wgauss}) is still
unspecified. It would be interesting, for example, to know which value of $\sigma$
minimizes the entropy $S_2[\ov W]$, within the bounds given by
Eq. (\ref{s_smooth2}). We have not been able to obtain a general result, but some
indication can be obtained from the following example. Let us suppose that
the function $W$ to be smoothed is also a Gaussian, as in Eq. (\ref{wgauss}),
but with spatial variance $\mu$ instead of $\sigma$.
The smoothed Wigner function is then
\be
\ov W(x,p) = W \conv G =
\frac{1}{2\pi\Sigma_x \Sigma_p} \exp\left(-\frac{x^2}{2\Sigma_x^2}
-\frac{p^2}{2\Sigma_p^2} \right)~,
\label{wgauss2}
\ee
with
\[
\Sigma_x^2 = \sigma^2 + \mu^2~~;~~~
\Sigma_p^2 = \frac{\hbar^2}{4}\left({1\over \sigma^2} + {1\over \mu^2}\right)~.
\]
The information corresponding to $\ov W$ is
\be
I[\ov W] = 2\pi\hbar \int \ov W^2 dx dp = \frac{\hbar}{2\Sigma_x \Sigma_p}~.
\ee
After some algebra, one obtains the following expression
\be
I[\ov W] = I(z) =\frac{z}{1+z^2}~,
\ee
where $z=\sigma/\mu$. The function $I(z)$ attains its maximum for $z=1$, i.e.
when $\sigma=\mu$, and the kernel has the same variance as the Wigner function to be
smoothed. In this case, $S_2[\ov W] = 1/2$, which represents the lower bound
of Eq. (\ref{s_smooth2}). We could conjecture, although we do not have
a formal proof, that this is the general result :
the minimum entropy increase due to smoothing with a Gaussian kernel
is attained when the width of the kernel is close to the width of the function
to be smoothed.

Another interesting example is provided by the harmonic
oscillator, whose Hamiltonian is
\be
H(x,p) = \frac{p^2}{2m} + m\omega^2\frac{x^2}{2}.
\label{hamilt}
\ee
The eigenstates can be expressed in terms of Hermite polynomials $H_n(\xi)$
($H_0=1,~ H_1=2\xi,~ H_2=4\xi^2-2,...$)
\be
\psi_n(x)=(2^n n!)^{-1/2} \left(\frac{m\omega}{\hbar\pi}\right)^{1/4}
\exp\left(-\frac{m\omega x^2}{2\hbar}\right) H_n(x\sqrt{m\omega/\hbar}).
\ee
The corresponding Wigner functions are \cite{tat}
\be
W_n(x,p) = \frac{(-1)^n}{\pi\hbar}\exp\left(-\frac{2H}{\hbar\omega}\right)
L_n\left(\frac{4H}{\hbar\omega}\right)~,
\ee
where $H(x,p)$ is the Hamiltonian, and the $L_n(\xi)$ are Legendre polynomials
($L_0=1,~ L_1=1-\xi,~ L_2=1-2\xi+\xi^2/2,...$). We now smooth such Wigner functions
with a Gaussian kernel, and find \cite{tat}
\be
\ov W_n(x,p) = (2\pi\hbar n!)^{-1} (H/\hbar\omega)^n \exp(-H/\hbar\omega)~.
\ee
Note that this relatively simple result for $\ov W_n$ is obtained only in the case
when the square variance of the smoothing kernel [see Eq. (\ref{wgauss})] is
$\sigma^2 = \hbar/2m\omega$; in all other cases the smoothed Wigner function
is not a function of the energy only. We are now in a position to compute the
information $I[\ov W_n] \equiv \ov I_n = 2\pi\hbar\int \ov W_n^2 dx dp$.
Let us first change to
polar coordinates $(r,\theta)$ in the phase space
\be
\frac{p^2}{m} + m\omega^2x^2 =  \hbar \omega r^2~,~~dx dp = \hbar r dr d\theta~.
\ee
One obtains, after integration over $\theta$
\be
\ov I_n = (n!)^{-2} \int_0^\infty (r^2/2)^{2n} \exp(-r^2) r dr~,
\ee
and finally, changing variable again $z=r^2/2$
\be
\ov I_n = (n!)^{-2} \int_0^\infty z^{2n} \exp(-2z) dz
= \frac{(2n)!}{2^{2n+1} (n!)^2}~.
\label{iharmo}
\ee
We first note that $\ov I_0 = 1/2$, in agreement with previous results, since
the ground state of the harmonic oscillator is a Gaussian, and we are smoothing
with another Gaussian of identical width. It can also be shown that $\ov I_n$
is a decreasing function of $n$. The asymptotic expansion (for $n \gg 1$) is
obtained by taking the logarithm of Eq. (\ref{iharmo})
and making use of Stirling's formula
\[
\ln X! \sim X\ln X -X + {1\over 2}\ln X~~~~~(X \gg 1)~,
\]
which yields
\be
\ov I_n \sim n^{-1/2}~.
\ee
In terms of the entropy, we have in summary
\be
\begin{array}{ll}
S_2[\ov W_0] = 1/2 \\
S_2[\ov W_{n+1}] > S_2[\ov W_{n}] \\
\lim_{n \to \infty}S_2[\ov W_{n}]  = 1~.
\end{array}
\ee
The latter results means that the entropy increase is larger when
smoothing a semi-classical state. Asymptotically, the entropy of the
smoothed Wigner function approaches unity. On the other hand, when
smoothing a `fully quantum' state (i.e. a state with small quantum
numbers), the entropy increase is moderate. Although these results were
obtained for the special case of the harmonic oscillator, we are confident
that they remain qualitatively correct for other
(classically integrable) Hamiltonians.

{\large{\bf VI. Discussion}}

In this paper we have presented several results related to
a new definition of quantum entropy, denoted $S_2$. Although it
has already been used in the past in the framework of the density
matrix formalism, such entropy becomes particularly interesting when
applied to Wigner functions. It is then possible to show that $S_2$
possesses a number of interesting properties $-$ most importantly, for example,
it is an invariant for the Wigner equation, which governs the evolution
of Wigner functions. $S_2$ is related to the Tsallis entropy, although the
latter is usually defined for a discrete set of probabilities, rather than
for a continuous distribution.
An advantage of this entropy, compared to the quantum Von Neumann entropy, is
that the Wigner function is all that one needs to compute $S_2$. No knowledge
of the density matrix is required, neither does it need to be
diagonalized, as is the case for the Von Neumann entropy.

The standard properties of entropy (concavity, additivity, sub-additivity) have
been examined. This has revealed some interesting facts, which would require
further investigations. For instance, it has been proven that $S_2$
(unlike ordinary entropy) behaves like a
probabilty with respect to additivity properties, which is also consistent with
the normalization $0 \le S_2 \le 1$.
Secondly, the analysis of the canonical ensemble has enabled us to derive
a Wigner function $W_{\rm{eq}}$ that maximizes the entropy under certain constraints.
$W_{\rm{eq}}$ turns out to be both a function of the energy alone and a
stationary solution of the Wigner equation. The relevance of $W_{\rm{eq}}$ is
still unclear, but one could reasonably conjecture that it plays a role in
some relaxation processes. Numerical experiments could clarify this point.

An "unpleasant" property of $S_2$ is that, keeping the Wigner function fixed, and
letting Planck's constant go to zero, one obtains $S_2=1$. Thus it would seem
that all classical states have unit entropy. The point is that this is not the
correct procedure to obtain a classical state: indeed, if the original Wigner function
is negative somewhere, we would obtain a classical state with a non-positive
probability distribution, which is of course meaningless. The correct procedure
is instead to smooth the Wigner function $W$ with an appropriate kernel, which must
also be a Wigner function in order to ensure positivity. A crucial point, however,
is that the smoothed Wigner function $\ov W$ should be itself an {\it admissible}
quantum state, i.e. one that can be described by a density matrix with non-negative
eigenvalues. We have been able to prove that, when smoothing with a minimum
uncertainty Gaussian packet, the result is always admissible $-$ although this is
{\it not} necessarily the case when smoothing with another Wigner function.
This is, to our knowledge, the first rigorous argument showing that Gaussian
smoothing possesses some privileged status.

It has also been proven that smoothing increases the entropy: in particular, when smoothing
a pure state with a Gaussian kernel, one has $S_2[\ov W] \ge 1/2$. It would be interesting
to know how to minimize $S_2[\ov W]$. This could be done by varying the width $\sigma$
of the Gaussian kernel, which is still a free parameter. Although we are not able to
derive a rigorous risult, we have conjectured (and shown explicitly on a particular
example) that $S_2[\ov W]$ is minimum when the width of the Gaussian kernel is close
to the width of the Wigner function to be smoothed. This would not be unreasonable
from the information point of view : it would mean that we can minimize the entropy increase
if we have some prior knowledge of the function to be smoothed.

As a further example, we have computed the entropy of the (smoothed) stationary states
of the harmonic oscillator. It was shown that  $S_2$ increases with quantum number,
therefore semi-classical states yield a larger entropy increase.
Again, we have conjectured that this behavior is universal (at least for confining
and classically integrable Hamiltonians),
and not specific to the harmonic oscillator. We are rather confident that
our conjecture is correct since the larger entropy increase for semi-classical states
is mainly due to the fact that their Wigner function displays short-wavelength
oscillations in the phase space, which are easily erased by the smoothing procedure.

It would be interesting to know how the previous results generalizes to
classically non-integrable Hamiltonians. For the harmonic oscillator, it was found
that the information of the smoothed stationary states behaves as
$\ov I_n \sim n^{-1/2}$. Although the exponent $-1/2$ might be specific
to the harmonic oscillator, a polynomial law may be universal for
the class of integrable Hamilationans. On the other hand, one could conjecture that,
for non-integrable Hamiltonians, the decrease is faster, perhaps exponential.

From the physical point of view, this result
means that semi-classical states are highly unstable
under generic perturbations (amongst which smoothing is a relevant example).
This is reminiscent of the so-called `predictability sieve', a concept introduced by
W.H. Zurek and co-workers \cite{zhp} in the more general framework of decoherence
\cite{decohere1,decohere2}. Zurek et al. \cite{zhp}
construct a model for the interaction of a quantum system
with an environment at thermodynamic equilibrium,
and compute the rate at which initially pure states
deteriorate into mixtures by coupling with the environment. This process
is known as decoherence. Subsequently,
they look for the set of states which are least prone to deterioration,
and find that such states are those which yield
the minimum entropy increase. By estimating the entropy production, they
obtain that the minimum-entropy increase is attained for the ground state
of the harmonic oscillator, i.e. a minimum uncertainty Gaussian wavepacket.
This coincides with our results of Sec. V.

The main difference from our approach is that
W.H. Zurek and co-workers \cite{zhp} analyze a {\it dynamical} situation,
while in our case
the entropy-producing effect is the smoothing, which is a static process.
Since both cases appear to give the same result, it is reasonable to
conjecture that smoothing may represent a (simplified) model for the
interaction of a quantum system with an open environment. The price to pay for
our approach is that we do not have a first-principle based derivation of such an
interaction. The advantage is that the model is simple enough to obtain a number
of rigorous results.

These considerations may shed some new light
on the semi-classical limit. We distinguish two kinds of pure quantum states:
fully quantum (FQ) states $W_{\rm FQ}$ (with low quantum numbers), and semi-classical
(SC) states $W_{\rm SC}$ (with
large quantum numbers). For both $S_2=0$, i.e. they contain the same amount
of information. However, after the smoothing, one obtains $S_2[\ov W_{\rm FQ}] \simeq 1/2$
and $S_2[\ov W_{\rm SC}] \to 1$, i.e. the smoothed FQ state
contains {\it more} information than the smoothed SC state.
In other words, although both original states contain the same information,
this is of different `quality': robust for the FQ state, and highly prone
to deterioration for the SC state. It is not surprising, therefore, that coupling
to an environment has the effect of erasing such information less easily
in the former case than in the latter.
These results could open new avenues for further research, particularly
with computer experiments \cite{num}, to investigate the dynamical behavior of the
entropy defined in this paper.

\newpage
{\large{\bf Appendix}}

We want to prove the result of Eq. (\ref{counter}).
Let us use the identity
\[
\int f(x,p)~ g(x,p)dx dp = \frac{1}{4\pi^2} \int f(k,\lambda)
g^{\conv}(k,\lambda)dk d\lambda~,
\]
with $f=\ov W$ and $g=W$.
Since $\ov W(x,p) = W \conv W$, the double Fourier
transform of $\ov W$ is $W^2(k,\lambda)$. In addition, it will turn out
that $W(k,\lambda)$ is real for the case under consideration here.
Therefore, by making use
of the previous identity, the left-hand side of Eq. (\ref{counter}) becomes
\[
\int \ov W(x,p) W(x,p) dx dp = \frac{1}{4\pi^2} \int W^3(k,\lambda)
dk d\lambda~.
\]
The double Fourier transform $W(k,\lambda)$ is given by Eq. (\ref{wigfour}).
For our example, the wavefunction is the one of Eq. (\ref{example}),
and we obtain
\[
W(k,\lambda)= 4\sqrt{2/\pi} \exp(-\lambda^2\hbar^2/2) \int \left( x^2
-\frac{\lambda^2\hbar^2}{4} \right) \exp(-2x^2) \exp(-ikx) dx~.
\]
Now, by using the following integrals
\[
\int \exp(-2x^2) \exp(-ikx) dx = \sqrt \frac{\pi}{2}~ \exp(-k^2/8)
\]
\[
\int x^2 \exp(-2x^2) \exp(-ikx) dx =
\sqrt\frac{\pi}{8}~(1-\frac{k^2}{4})~\exp(-k^2/8)~,
\]
we obtain, after some straightforward algebra
\[
W(k,\lambda)= \left(1-\frac{k^2}{4}-\lambda^2\hbar^2 \right)
\exp\left(-\frac{\lambda^2\hbar^2}{2}-\frac{k^2}{8}\right)~.
\]
We are now ready to compute the integral $\int W^3 dk d\lambda$.
Let us change integration variables $(k,\lambda) \to (r,\varphi)$
\[
r^2 = \frac{k^2}{4} + \lambda^2\hbar^2~~;~~ r dr d\varphi =
\frac{\hbar}{2} d\lambda dk~.
\]
After integration over $\varphi$, one obtains
\[
\int_{-\infty}^\infty \int_{-\infty}^\infty
W^3 dk d\lambda = \frac{4\pi}{\hbar}~\int_0^\infty  (1-r^2)^3 \exp
(-\frac{3}{2} r^2)~r dr~.
\]
Changing the integration variable to $y=r^2$ and using integrals of the type
\[
\int_0^\infty y^n \exp(-ay) dy = \frac{n!}{a^{n+1}}~,
\]
it is obtained
\[
\int_{-\infty}^\infty \int_{-\infty}^\infty
W^3 dk d\lambda = \frac{2\pi}{\hbar}~\int_0^\infty  (1-y)^3 \exp
(-\frac{3}{2} y)~dy = -\frac{4\pi}{27\hbar}~,
\]
which, once divided by $4\pi^2$,
yields the result of Eq. (\ref{counter}).

\end{document}